# Spatial-Temporal Inference of Urban Traffic Emissions Based on Taxi Trajectories and Multi-Source Urban Data

Jielun Liu[a], Ke Han[b], Xiqun (Michael) Chen[c,*], Ghim Ping Ong[a]

[a] Department of Civil & Environmental Engineering, National University of Singapore, 117576, Singapore
[b] Center for Transport Studies, Department of Civil and Environmental Engineering, Imperial College London, London SW7 2BU, UK
[c] College of Civil Engineering and Architecture, Zhejiang University, Hangzhou 310058, China

**Abstract**

Vehicle trajectory data collected via GPS-enabled devices have played increasingly important roles in estimating network-wide traffic, given their broad spatial-temporal coverage and representativeness of traffic dynamics. This paper exploits taxi GPS data, license plate recognition (LPR) data and geographical information for reconstructing the spatial and temporal patterns of urban traffic emissions. Vehicle emission factor models are employed to estimate emissions based on taxi trajectories. The estimated emissions are then mapped to spatial grids of urban areas to account for spatial heterogeneity. To extrapolate emissions from the taxi fleet to the whole vehicle population, we use Gaussian process regression models supported by geographical features to estimate the spatially heterogeneous traffic volume and fleet composition. Unlike previous studies, this paper utilizes the taxi GPS data and LPR data to disaggregate vehicle and emission characteristics through space and time in a large-scale urban network. The results of a case study in Hangzhou, China, reveal high-resolution spatio-temporal patterns of traffic flows and emissions, and identify emission hotspots at different locations. This study provides an accessible means of inferring the environmental impact of urban traffic with multi-source data that are now widely available in urban areas.

# 1. INTRODUCTION

The rapid global economic growth and urban development have led to aggressive urbanization and motorization in many developing countries since the 1980s, which has contributed significantly to the amount of vehicle-derived air pollution. The number of registered automobiles in China has skyrocketed from 11 million in 1996 to 163 million in 2016 (National Bureau of Statistics of China, 2017). The increased vehicle usage significantly contributes to greenhouse gas emissions and urban air pollution. Large cities, especially those in developing countries, are suffering from air pollution and induced adverse health impacts. The World Health Organization (2014a, 2014b) reported that urban air quality has become a focus of public concern, estimating that ambient air pollution caused approximately 3.7 million premature deaths every year. Moreover, atmospheric pollutants can easily spread to a large scale, influencing more people than one could expect.

In recent years, there has been an increasing number of institutions and researchers focusing their research on the issue of vehicle emission and urban air pollution (Gühnemann et al., 2004; Zhang et al., 2008; WTO, 2014a, 2014b; Tan et al., 2015; Nyhan et al., 2016; Ma et al., 2018). One way to investigate the environmental impact of road traffic is to derive emission inventories, which are used to estimate the total traffic emissions in a specified area, global (Butler et al., 2008; Wang et al., 2014) or regional (Zhang et al., 2008), within a specified time span (usually an annual year). An emission inventory can provide macroscopic emission information on urban cities. For example, a region-based emission inventory can reveal spatial patterns of urban vehicle emission. Emission inventories of several large cities have been put forward and discussed in the literature, which can help the government make proper policies such as vehicle restrictions to reduce emissions.

Several approaches have been developed to analyze vehicle emissions and urban air pollution, and to assess the impact of traffic on air pollution. For instance, Wang et al. (2010) proposed a vehicle emission model to study the vehicular emission trend in China based on vehicle mileage traveled (VMT) and emission factors (EF). Stead (1999) adopted the EF (average kilometer ratio) to calculate vehicle emission. Emission models such as MOVES (Koupal et al., 2003), COPERT (Nikoleris et al., 2011), and the IVE model (Guom et al., 2007; Zhang et al., 2008) were developed and adapted based on local vehicle information databases (e.g., vehicle types and fuel types). MOVES was developed by the U.S. Environmental Protection Agency to estimate emissions (including greenhouse gas and toxic pollutants) from on/off-road mobile sources. It considers several mobile emission processes including running exhaust, brake wear, tire wear, and running loss. COPERT is a commonly used emission model in Europe. This model utilizes a large amount of experimental data to determine emission parameters from road transportation and to obtain emission inventories. The IVE model adopts the vehicle specific power (VSP) and engine stress (ES) as inputs to calculate EF. However, emission models like MOVES and IVE use inherent databases, the parameters need to be carefully calibrated using local information while applying these models to analyze different areas. Recently, Jamshidnejad et al. (2017) proposed a general framework combining micro and macro emission models to estimate vehicular emissions which integrated traffic flow.

There are several approaches that rely on drive cycle characteristics (such as speed and acceleration) and vehicle types to determine EFs. These approaches are typically based on local vehicle emission characteristics as determined from laboratory or on-road measurements and include the chassis dynamometer measure system (Pelkmans and Debal, 2006), the portable emission measurement system (PEMS) (O'Driscoll et al., 2016), and the remote sensing detection (RES) (Wang et al., 2011). PEMS and RES are usually applied to determine emissions characteristics (Wang et al., 2011) or compare results against different emission models (Ekstrom et al., 2004). Such emission models usually estimate the total emissions in urban areas, while several other studies in the literature have applied similar methods to estimate emissions and their spatial distributions. For example, Luo et al. (2017) analyzed the spatial-temporal features of taxis' emissions in Shanghai, although they did not estimate the emission of the entire vehicle population. Nyhan et al. (2016) utilized taxi GPS data and microscopic emission models to predict the spatial-temporal distribution of vehicle emissions in Singapore. In their study, the total emissions from the entire vehicle population were extrapolated from taxi-based emission, using loop detector counts and static fleet composition information in the country.

To investigate the dynamic nature of vehicle emission, researchers have also developed various pollutant diffusion models to study urban vehicle pollution under different circumstances. For instance, the STREET model considered roads surrounded by tall buildings (Johnson et al., 1973). The Gaussian dispersion model was found to be suitable for the open area (Benson, 1992). Land-use-based regression models were developed to predict intra-urban air pollution (Briggs et al., 1997; Clougherty et al., 2008). Csikós et al. (2015) proposed combined simple diffusion models to investigate the dispersion of motorway traffic.

Despite many approaches developed in the literature to model vehicle emissions, the effective estimation of traffic emissions in an urban environment remains a significant challenge. This is largely attributed to the insufficient coverage of relevant sensors and lack of accurate information for validation. Combining multi-source data provides a viable solution to tackle this challenge. Nyhan et al. (2016) utilized taxi GPS data and loop detector data to predict the spatial-temporal distribution of vehicle emissions in Singapore. Shang et al. (2014) utilized the taxi trajectory data and geographical information to infer vehicular emissions in Beijing. Floating car data (FCD) were used to reveal how emissions affected residents in Berlin (Gühnemann et al., 2004). An artificial neural network model was adopted to identify the total taxi air pollution emissions based on taxi data (Zeng et al., 2007). Xu et al. (2017) utilized mobile phone data to obtain origin-destination matrices, with which some traffic patterns and performance indices were acquired, and applied the random forest method to reveal the impact of traffic on air quality predictability in Beijing. Mo et al. (2017) proposed a speed estimation model using license plate recognition (LPR) data which could support for calculating vehicular emission in micro-emission model. Nocera et al. (2018) estimated carbon emissions from road transport using incomplete traffic information collected by flow estimators.

Most of the aforementioned studies focused on emissions from probes (e.g., taxis) instead of the entire vehicle population. The few exceptions (e.g., Nyhan et al., 2016) tend to extrapolate from probe emissions based on static and aggregated fleet information without consideration of spatial and temporal heterogeneity of such information. To bridge the gap, this paper proposes a framework for inferring the spatio-temporal traffic volume and fleet composition with reasonable accuracy and granularity, which, when combined with taxi trajectory data, yields a reliable estimation of total traffic emissions on a city scale. This is achieved by fusing multi-source and heterogeneous urban datasets (e.g., LPR data, Point of Interests data, and urban road network data) using Gaussian process regression (GPR) models. As we subsequently show in the case study, consideration of these additional features offers greater accuracy and granularity of the spatial and temporal dynamics of vehicle emissions, compared to existing methods (Nyhan et al., 2016).

The emission profiles of taxis are estimated with the vehicle emission model and then mapped to the spatial grids of an urban area to account for spatial heterogeneity. To extrapolate emissions from the taxi fleet to the entire vehicle population, we use GPR models supported by a variety of geographical features to estimate time-varying and spatially heterogeneous traffic volume and fleet composition, which informs the coefficients of extrapolation. A case study in the city of Hangzhou reveals spatio-temporal traffic emission patterns and identifies several emission hotspots.

The main contributions of this paper are summarized as follows:

(a) We propose a practical method to infer the vehicle volume and fleet information on a city scale, by combining multi-source urban data with statistical learning techniques. This bridges the aforementioned gap of the insufficient penetration of probe sensors or partial knowledge on the vehicle fleet composition that is difficult to measure in a large-scale urban network.

(b) This paper presents a viable way of recovering the carbon footprint of urban traffic with reasonable spatio-temporal granularity and accuracy, by utilizing urban datasets that are nowadays widely accessible.

(c) The proposed methodology has a high transferability to other studies that aim to reveal urban dynamics based on sparse or partial observations, and contributes to the broader scientific inquiry of data transferability.

The remainder of the paper is organized as follows: Section 2 proposes the methodological approach to the spatial-temporal analysis of vehicle emissions. Section 3 presents the emission pattern results of the city-wide case study in Hangzhou, China using the approach proposed in Section 2. Finally, Section 4 concludes the paper and provides an outlook on future research.

## 2. METHODOLOGY

In this section, we propose the framework for estimating vehicle emission in an urban network by utilizing both taxi GPS data and LPR camera data, as shown in Figure 1. Specifically, we first adopt emission models to calculate the taxi emissions based on probe data (Section 2.1). Then, we allocate emission along each trajectory to different spatial grids (Section 2.2). Third, we use GPR models combining taxi GPS data, information provided by LPR cameras, and geographic information to estimate the spatially heterogeneous vehicle count and fleet composition, which are related to the entire vehicle population (Section 2.3).

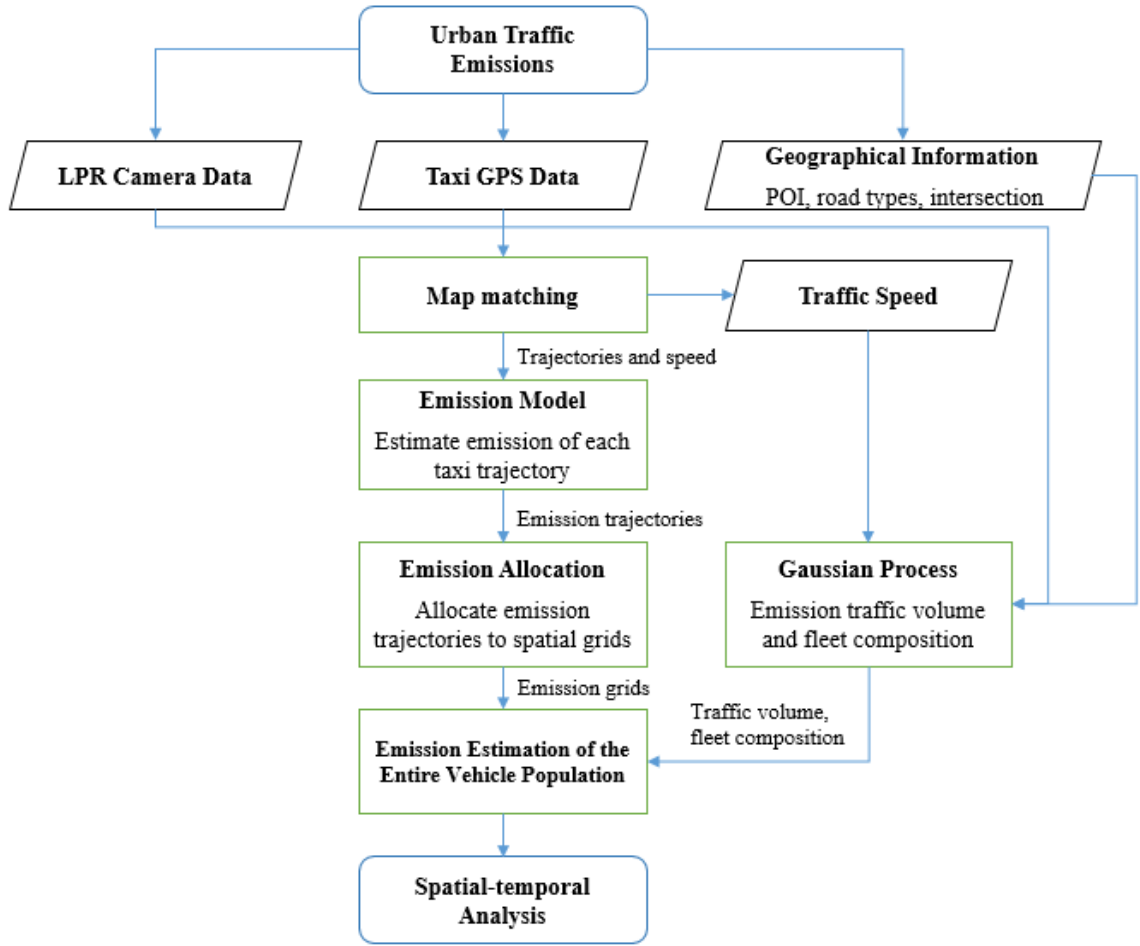

Figure 1. The framework for the spatial-temporal inference of urban traffic emissions.

The following notations are introduced.

| | | |
|---|---|---|
| $d$ | = | travel distance |
| $Ec(i,t)$ | = | extrapolation coefficient in region $i$ at time $t$ |
| $EF$ | = | modified emission factor |
| $EF_0$ | = | emission factor |
| $EF_H(i,t)$ | = | emission factor of heavy vehicles in region $i$ at time $t$ |
| $EF_k(i,t)$ | = | emission factor of vehicle type $k$ in region $i$ at time $t$ |
| $EF_L(i,t)$ | = | emission factor of light vehicles in region $i$ at time $t$ |
| $EF_{taxi}(i,t)$ | = | emission factor of taxis in region $i$ at time $t$ |
| $EF_{relative}$ | = | relative emission factor |
| $N_H(i,t)$ | = | number of heavy vehicles in region $i$ at time $t$ |
| $N_k(i,t)$ | = | number of vehicles of vehicle type $k$ in region $i$ at time $t$ |
| $N_L(i,t)$ | = | number of light vehicles in region $i$ at time $t$ |
| $N_{taxi}(i,t)$ | = | number of taxis in region $i$ at time $t$ |
| $time_{i,j}$ | = | timestamp of taxi $i$ at sampling point $j$ |
| $v_{i,j}$ | = | average speed of taxi $i$ on segment $j$ (between sampling point $j$ and $j+1$) |

## 2.1. Emission Model

In this section, we introduce the emission model adopted to estimate taxi emissions. The emission model is given by

$$Emission = EF \cdot d = EF_0 \cdot EF_{relative} \cdot d \qquad (1)$$

where *Emission* can be CO, HC, or $NO_x$; $EF_{relative}$ is the relative emission factor used to modify emission factor $EF_0$ ($EF_0$ can be seen as the basic emission factor, and $EF$ varies with speed); $d$ is the travel distance.

We adopt the vehicle emission factor models developed in (Wang et al., 2014; Zhang et al., 2012, 2014). The vehicle emission factor model uses laboratory- or on-road measured drive cycles, traffic patterns, and emission control levels in the determination of EF. We use the average speed calculated between two consecutive timestamps to approximate the instantaneous speed, given by

$$v_{i,j} = \frac{distance(point_{i,j}, point_{i,j+1})}{time_{i,j+1} - time_{i,j}} \qquad (2)$$

where $v_{i,j}$ is the average speed of taxi *i* on segment *j* (between sampling points *j* and *j+1*); $time_{i,j}$ is the timestamp of taxi *i* at sampling point *j*; $distance(point_{i,j}, point_{i,j+1})$ is the distance between two sampling points of taxi *i*.

There are many possible trajectories. The real distance between two adjacent points could not be calculated directly if the actual trajectory remains unknown. Using the straight-line distance between two consecutive points to estimate the distance could possibly result in an underestimation of distance traveled. In this study, map matching algorithms are employed to provide the actual or most possible trajectory between the two adjacent points. Generally, there are four kinds of map matching algorithms based on the additional information they use, namely geometric (Greenfeld, 2002), probabilistic (Newson and Krumm, 2017; Wan et al., 2016), topological (Velaga et al., 2009), and mapping-to-cell techniques (He et al., 2017). In term of different sampling frequencies, there are map matching algorithms preferable for low-frequency data (Quddus and Washington, 2015; Chen et al., 2014) and relatively high-frequency data (Newson and Krumm, 2017). The sampling frequency of taxi GPS data is usually within one minute and due to the high performance, we adopt the map matching algorithm proposed by Newson and Krumm (2017). With map matching algorithms, $distance(point_{i,j}, point_{i,j+1})$ becomes the travel distance of the mapping trajectory between two sampling points of taxi *i*.

To present dynamic emission patterns, we divide one day into 144 episodes with a time interval of 10 min and calculate the total emission from taxis for every episode. Note that the number of episodes is user-defined. In our study, we use the episode with 10 min as it is adequate to reveal the short-term traffic conditions. This will return a set of emission data with spatial references, which are then utilized to allocate emissions to spatial grids and analyze their spatial and temporal patterns. Besides, this emission model is utilized to determine EF of taxis and other types of vehicles, which will be discussed in detail in Section 2.3.

## 2.2. Taxi Emission Allocation

The geographic referencing of vehicle trajectories or emissions can be done via map matching, either on road segments (Nyhan et al., 2016) or on user-defined spatial grids (Luo et al., 2017). This paper adopts the latter method. The dimension of a grid may influence the results, as a smaller grid may suffer from noise while a larger grid may miss some spatial patterns. We divide the urban area of interest into grids with a dimension of 100 m × 100 m per grid (also called region in this study), and allocate the vehicle-based emissions to every grid. Estimation of emission from every segment of the taxi trajectory (between every two contiguous sampling points) can provide several taxi emission trajectories. Then, the taxi emissions along trajectories can be allocated to these grids. Taking Figure 2 as an example, A and D represent two consecutive sampling points, and the emission computed between A and D are uniformly distributed to the three sub-segments: AB, BC, and CD. In other words, the amount of emissions that Grids 1 through 3 receive from this particular trajectory is proportional to the lengths of AB, BC, and CD, respectively:

$$\begin{cases} Emission_{AB} = \dfrac{AB}{AD} \times Emission_{AD} \\ Emission_{BC} = \dfrac{BC}{AD} \times Emission_{AD} \\ Emission_{CD} = \dfrac{CD}{AD} \times Emission_{AD} \end{cases} \quad (3)$$

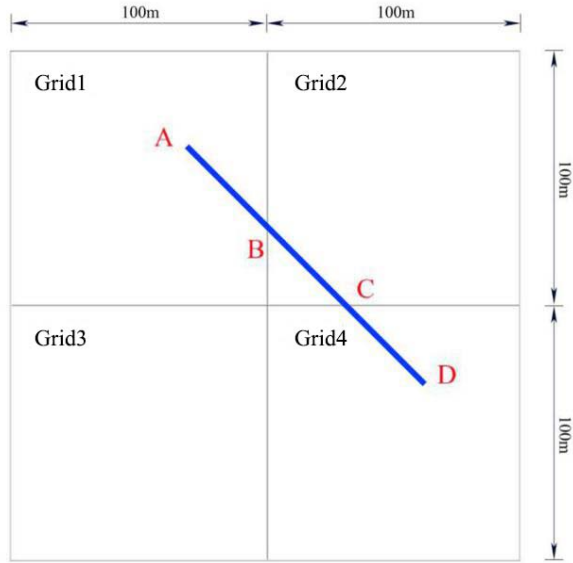

Figure 2. Allocation of vehicle emission trajectory to grids.

Since each taxi trajectory comprises a series of line segments, the abovementioned method allows us to calculate the emission contribution of individual taxis to the corresponding grids. Then we can estimate the grid-based emission by summing up all such allocated emissions to the object grid.

## 2.3. Emission Estimation of the Whole Vehicle Population

The previous two steps estimate the grid-based emissions generated from taxi trips, thereby presenting the spatial distribution of taxi emissions. To extend the emission estimation from taxi to the entire vehicle fleet, the following three factors should be considered: vehicle number (or flow), fleet composition, and emission factors that vary by vehicle type. Data extracted from LPR cameras are used to estimate the total number of moving vehicles and the fleet composition, at discrete locations. As these three factors can vary in time and space, there is a need to explicitly consider their spatio-temporal variation for a more accurate depiction of the emission profile within the traffic network. This is in contrast to other simplified methods of extrapolating taxi emissions for modeling general multi-class traffic emission in the literature (Nyhan et al., 2016; Luo et al., 2017). To reveal the fine-grained spatial patterns, we further mine geographical information on the study area.

*2.3.1 Vehicle Count*

We consider the vehicle tag information captured by LPR cameras, which consist of the camera ID, camera location (including road segment, direction, and coordinates), vehicle license, and vehicle type. Initially, we attempted to estimate vehicle count in each spatial grid via a weighted vote from the nearby LPR cameras (e.g., the *k*-nearest neighbor algorithm). However, due to the sparse and uneven distribution of LPR cameras in the road network, it is difficult to determine the exact spatial distribution of the traffic volume and fleet composition. While studies like Nyhan et al. (2016) utilized static and aggregate information on the fleet composition for estimating the extrapolation coefficient, such information is refined in this paper through a spatio-temporal reconstruction of the fleet information, which allows a more accurate estimation of emissions.

Following Section 2.2, a spatial discretization of the study area (into 100 m × 100 m square grids or regions) is performed. The traffic volume and fleet composition in each grid need to be inferred. To account for the diversity of spatial features pertinent to traffic volume and fleet composition, we utilize additional geographical information including Points of Interests (POI) of each gird mined from Gaode Map API, road types and number of signalized intersections from Open Street Map (OSM). Each of these features is elaborated as follows:

- POIs are categorized into several types, such as transportation, entertainment, companies and offices, residences, education, tourist spots, restaurants, malls and shopping, hotel, banks and ATMs. In a given grid, each type of POI is counted and the POI features can be thus represented as a vector, such as [0, 1, 0, 3, 0, 0, 2, 0, 3, 0], which can be further fitted into a distribution. Two vectors that are close to each other under certain metric means that the two regions share similar POI features. The number of POIs in a region is a positive indicator of its vitality and bustlingness, and is closely related to its attraction of travel demands.

- Road types are categorized as primary, secondary, tertiary and others[*] (as categorized by OSM). Like POI, road type can be also represented with a vector, for instance, [0.203, 0, 0.096, 0.108] means that the total lengths of primary roads, secondary roads, tertiary roads and other types of roads in the region are respectively 0.203 km, 0 km, 0.096 km, and 0.108 km. Road type is considered a relevant feature as certain types of vehicles are only allowed to use certain road types. The flow capacities of different types of roads also vary, which contributes to the differences in traffic volume.

---

[*] In US and Europe, we tend to refer to highways/expressways, major arterials, minor arterials, collectors and distributors and local roads. In this study, we follow the categories given by OSM.

- The number (or spatial density) of signalized intersections in a grid is linked to the road type (e.g., urban expressway or arterial) as well as the surrounding built environment, which is envisaged to be related to traffic volume and fleet composition.

To incorporate the abovementioned spatial features, geostatistical algorithms such as the GPR – a widely used model in the domain of spatial analysis – is adopted in this paper. Gaussian process (GP) is a stochastic process assuming that any finite combination of samples has a multivariate normal distribution. GP is parameterized by a mean function and a covariance function:

$$\mathbf{f}(\mathbf{x}) \sim GP\left(\mu(\mathbf{x}), K(\mathbf{x}, \mathbf{x}^{\mathrm{T}})\right)$$

where $\mathbf{f}(\mathbf{x}) = [\text{vehicle count, fleet composition}]^{\mathrm{T}}$ denotes the vehicle count and fleet composition, which are estimated with separate GP models. $\mathbf{x} = [\text{location, POI, road types, no. of intersections, average speed, no. of taxis}]^{\mathrm{T}}$ is the input of GP models; $K$ is the kernel function. In this study, we apply the sequence exponential kernel. The parameters of the GPR model can be determined with the observed data, and then we can estimate the vehicle count and fleet composition in each region.

*2.3.2 Emission Factors*

We adopt the method developed in Section 2.1 to estimate emission factors of different vehicle types. However, speed should be estimated prior to calculating emission factors. Inspired by the research carried out by Aslam et al. (2012), which demonstrated that taxi GPS data could be used to infer general traffic patterns, we estimate the average speed in each grid every 10-minute episode based on taxi GPS data, and use the estimated speed to determine emission factors. It should be noted that when estimating the average speed in each grid, the zero-speed records of taxis will underestimate the actual speed and we only consider the speed of stationary vehicles if they remain stationary within 5 min to account for possible passenger waiting, boarding or alighting. As the speed undulates from different grids in different time periods, the emission factors not only present temporal patterns but also present spatial differences.

*2.3.3 Fleet Composition*

The fleet composition is estimated using the GPR model proposed in Section 2.3.1 for each episode. Since the emissions of taxis have already been estimated, we can compute the total emission from the entire vehicle population as follows:

$$Ec(i,t) = \frac{\sum N_k(i,t) \cdot EF_k(i,t)}{N_{taxi}(i,t) \cdot EF_{taxi}(i,t)} \tag{4}$$

where $Ec(i,t)$ is the extrapolation coefficient in region $i$ at time $t$; $N_k(i,t)$ represents the number of vehicles of type $k$; $N_{taxi}(i,t)$ represents the number of taxis; $EF_k(i,t)$ represents emission factor of vehicle type $k$; $EF_{taxi}(i,t)$ represents emission factor of taxis. It should be noted that $N_k(i,t)$ and $N_{taxi}(i,t)$ are vehicle counts with temporal patterns, while $EF_k(i,t)$ and $EF_{taxi}(i,t)$ are emission factors with both spatial and temporal patterns. $Ec(i,t)$ is calculated every time period (in this paper, it is calculated every 10 min).

The extrapolation coefficient is proposed to compute the total emission in each region. The total emission of the entire traffic population is calculated as emissions from taxis multiplied by the extrapolation coefficient. Note that the emission from taxis and the extrapolation coefficients vary spatially and temporally.

*2.3.4 Areas without Probe Data or LPR Data*

Due to the short time interval (e.g., 10 min) of each episode, it is likely that there are no probe data or LRP data in some areas, especially during the midnight when the traffic volume is quite low. For areas without probe data or LPR data, we adopt the following ways to process the data.

- Probe data: Check whether historical records exist in these areas. If there are records in historical data, we replace with a value given by the interpolation using the nearest recorded areas. Otherwise, the estimation will not be conducted in these areas.

- LPR data: Check whether historical records exist in these areas. If there are records in historical data, we replace with a value given by time series. Otherwise, we will not consider these areas as inputs in the GPR model. Such a vehicle count will be considered as an input in the GPR model if it passes the leave-one-out cross-validation.

## 3. CASE STUDY

### 3.1 Study Area and Data

In this section, the proposed method is applied to estimate the spatial and temporal patterns of vehicle emissions in Hangzhou, China. Hangzhou is the capital and most populous city of Zhejiang Province, China. The study area (shown in Figure 3) covers approximately 2,736 km$^2$.

The dataset used in this study involves 7,259 taxis and 271,117,329 data entries over a period of 9 days (June 22-30, 2015). The duration between two consecutive samples of GPS positioning varies between 5 s and 50 s, with the majority being 20-30 s, and a few up to 180 s because of data transferring errors. As shown in Table 1, the data typically contain the location of the taxi (longitude and latitude), time, taxi ID, operation status (occupied/vacant), instantaneous speed, and movement direction. The spatial discretization of the study area (Hangzhou) into 100 m × 100 m squares results in 480 × 570 grids. POI of these grids (totally 273,600 regions) is mined from Map API.

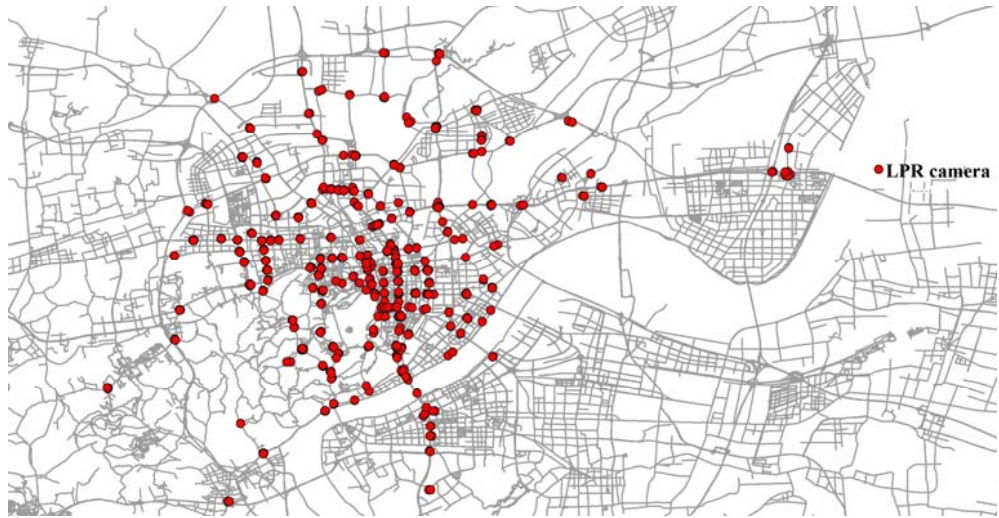

Figure 3. The study area and locations of LPR cameras.

Table 1. A sample of taxi GPS data

| Date | Time (h:min:s) | Taxi ID | Longitude | Latitude | Speed (km/h) | Direction (degree) | Status (1 occupied or 0 vacant) |
|---|---|---|---|---|---|---|---|
| 20150629 | 23:59:59 | 18338 | 118.254585 | 31.403702 | 63.3 | 140 | 1 |
| 20150629 | 23:59:55 | 13044 | 120.1437 | 30.338099 | 2.2 | 0 | 0 |
| 20150629 | 23:59:58 | 22320 | 120.16105 | 30.279967 | 0 | 135 | 0 |

It should be noted that missing values and invalid points due to data recording or transferring errors have been excluded in the following cases:

- Records with a missing ID or other information is eliminated;

- If a taxi doesn't move during a trip, this trip will be eliminated (status errors due to transferring errors or drivers' behavior);

- If the speed is larger than 130 km/h (which only takes up 0.002% of the whole dataset), this record will be regarded as abnormal records and thus eliminated;

- If a trip only contains one record, it will not be considered as a complete trip;

- If the time gap between two captured GPS signals is longer than 180 s, these two records will not be utilized in the emission-allocation process (as they may significantly influence the emission pattern along the trajectory).

After map matching, the actual/most possible trajectory between every two contiguous points is acquired, based on which the travel distance is calculated. Due to the data recording or transferring errors, data collected from about 6,700 taxis out of 7,259 were utilized for map matching.

The taxi emissions are determined by the model presented in Section 2.1. It should be noted that this model does not consider the case when the vehicle speed is below 5 km/h because idling preponderates in this stage, which has significant influences on determining EF. Details on the relative emission factors (Wu et al., 2012; Wang et al., 2014; Zhang et al., 2012, 2014) are shown in Table 2, where the emissions of CO, HC, and NOx are considered in the case study of Hangzhou. The study period is divided into 1,296 episodes of 10 min (9 days, and 144 episodes per day).

Table 2. Details on the relative emission factors

| Emission Type | $EF_0$ | $EF_{relative}$ |
|---|---|---|
| CO (light vehicle) | 0.86g/km | $EF_{relative} = 0.0006v_{i,j}^2 - 0.0627v_{i,j} + 2.2656$ |
| HC (light vehicle) | 0.08g/km | $EF_{relative} = 0.0006v_{i,j}^2 - 0.0762v_{i,j} + 2.7179$ |
| NOx (light vehicle) | 0.09g/km | $EF_{relative} = -0.443\ln(v_{i,j}) + 2.4102$ |
| CO (heavy vehicle) | 4.20g/km | $EF_{relative} = 8.857v_{i,j}^{-0.661}$ |
| HC (heavy vehicle) | 0.14g/km | $EF_{relative} = 12.093v_{i,j}^{-0.742}$ |
| NOx (heavy vehicle) | 7.50g/km | $EF_{relative} = 5.446v_{i,j}^{-0.50}$ |

Note: $EF_0$ and $EF_{relative}$ vary with the vehicle type.

Regarding the extrapolation of taxi emissions to the entire vehicle fleet, we utilize the vehicle tag information provided by 681 LPR cameras (located in 212 intersections or road segments) over the study period to determine the vehicle count, vehicle types and fleet composition in the road network. This LPR dataset contains 15,506,169 records. The raw LPR data consist of the LPR ID, LPR location (including road segment, direction, and coordinates), and vehicle type. Some sample LPR data are shown in Table 3. In AM/PM peak hours, the LPR cameras record about 220,000 vehicles in the study area. As shown in Figure 3, the LPR cameras are unevenly distributed in the network.

We obtain the operating pattern of taxis and other types of vehicles using the taxi GPS dataset and the LPR data. Since the taxi GPS data span the entire 24 hours per day, we recognize that a taxi is not operating if it does not move for more than 20 min. Figure 4 shows the number of operating taxis throughout the 24-hour horizon.

Table 3. A sample of LRP data used in this study

| Date | Period | Camera ID | Longitude | Latitude | Vehicle Type | Vehicle Count |
|---|---|---|---|---|---|---|
| 20150622 | 0:50am-1:00am | 310003000003 | 120.0939 | 30.2888 | 1 | 104 |
| 20150622 | 9:50am-10:00am | 310003000003 | 120.0939 | 30.2888 | 2 | 39 |
| 20150622 | 11:10am-11:20am | 310003000004 | 120.1204 | 30.3089 | 1 | 495 |

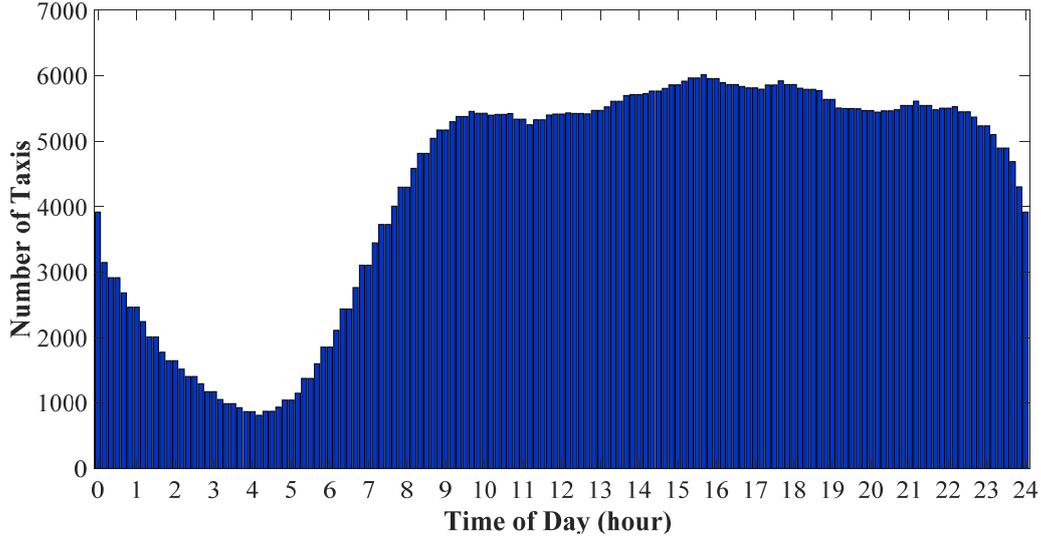

Figure 4. The number of taxis in service within a sample day (June 30, 2015, Tuesday).

Figure 5 shows the time-varying number of taxis, light vehicles, and heavy vehicles (observed by LPR cameras), which are the types of vehicles considered in this case study. It is clear that light vehicles take up the majority of the vehicle population in the road network. To calculate the extrapolation coefficient of each grid, Eq. (4) is changed into

$$Ec(i,t) = \frac{N_L(i,t) \cdot EF_L(i,t) + N_H(i,t) \cdot EF_H(i,t)}{N_{taxi}(i,t) \cdot EF_{taxi}(i,t)} \tag{6}$$

where $N_L(i,t)$, $N_H(i,t)$ and $N_{taxi}(i,t)$ are the numbers of light vehicles, heavy vehicles, and taxis in region $i$ at time $t$, respectively. $EF_L(i,t)$, $EF_H(i,t)$ and $EF_{taxi}(i,t)$ are emission factors of light vehicles, heavy vehicles, and taxis in region $i$ at time $t$, respectively.

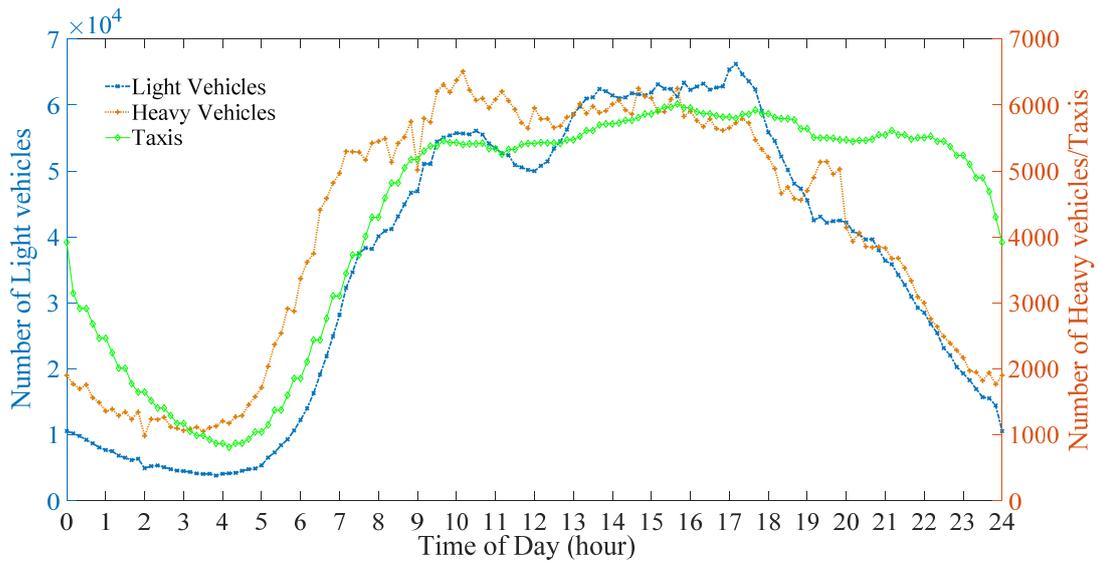

Figure 5. Observed dynamic vehicle composition from taxi data and LPR data by time of day.

## 3.2 Results

In this section, we first analyze the taxi emissions based on taxi GPS data, and then analyze the emission patterns of the entire vehicle population using the emission extrapolation method. The temporal patterns of taxi emissions (CO, HC, and $NO_x$) during the 9 days are shown in Figure 6. It is shown that the emission level is at the lowest during 4-5 AM, and increases continuously since 5 AM, reaching its peak at 3 PM. In addition, the emission patterns of taxis do not exhibit peaks during rush hours because the number of taxis in the road network does not have a sharp change in rush hours (due to the taxi drivers' shift work schedule).

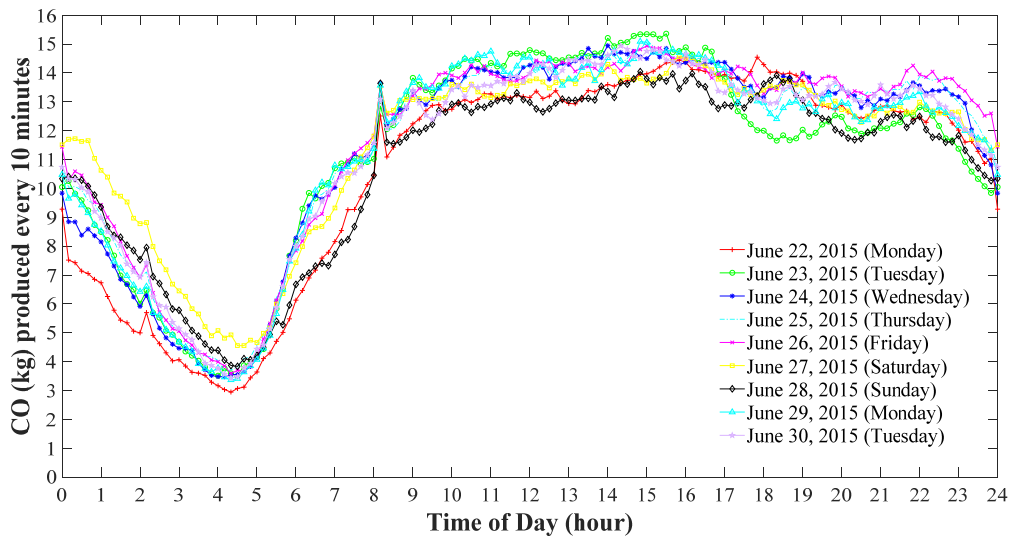

(a) CO emission

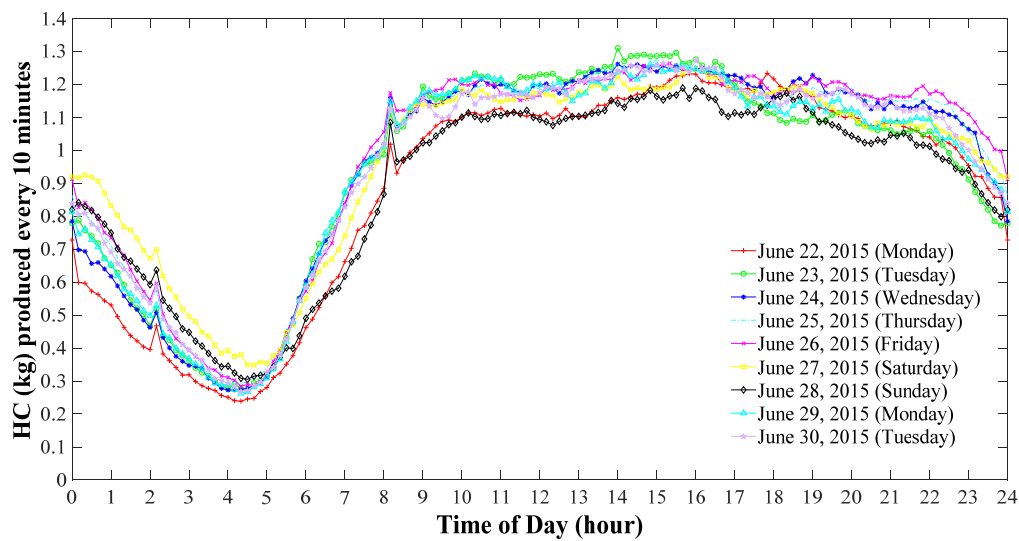

(b) HC emission

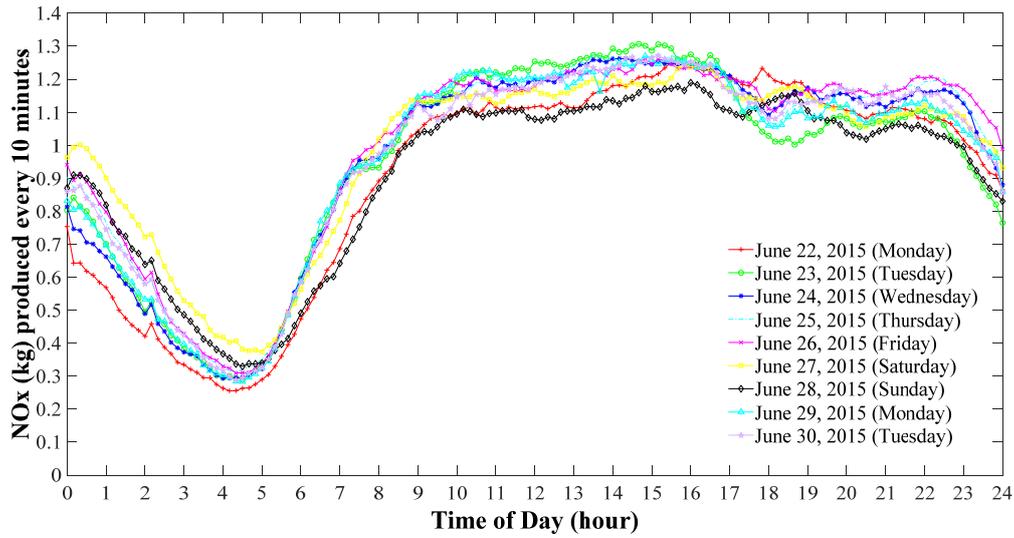

(c) NOx emission

Figure 6. Estimated emissions of taxis for different pollutants by the time of day.

Figure 7 shows the spatial distribution of vehicle counts estimated by the proposed GPR model (5:20-5:30 PM, June 22, 2016). It can be seen that the downtown area of Hangzhou experiences the heaviest traffic in terms of volume, while traffic away from the city center decreases. In addition, it is clear that the GPR model is capable of reconstructing spatially heterogeneous traffic volumes in the entire study area, based on limited traffic observations and additional spatial features illustrated in Section 2.3.1.

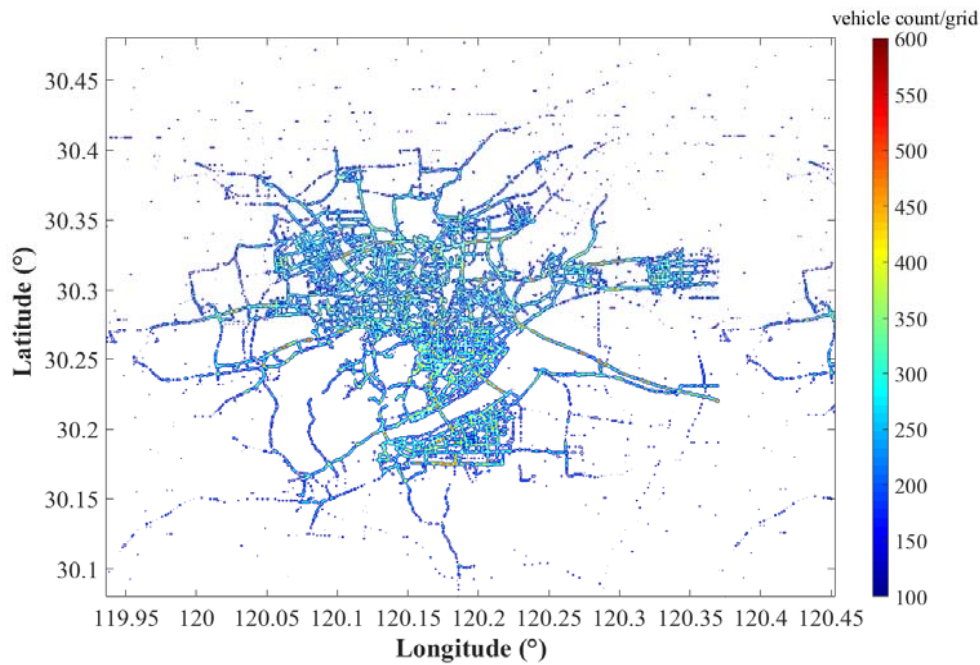

Figure 7. Spatially heterogeneous vehicle counts for a given 10-minute episode during a sample day (5:20-5:30 PM on June 23, 2015).

To understand the performance of the proposed GPR model, we compare it with two baseline models: Random forest, and GPR without additional spatial information (OSM data and POI information). Random forest uses the same input variables as GPR and the number of decision trees is set as 100 in this case. The dataset is split into a training subset (70%) and a test subset (30%). Other parameters are under fine-tuned. Table 4 compares the performances of the three methods of estimating the spatially heterogeneous vehicle counts (per minute per grid) and fleet composition (the ratio of light vehicles over the whole vehicle population).

Table 4. Estimation performance on the vehicle count and fleet composition

| Parameter | Estimation method | Performance measure | | |
|---|---|---|---|---|
| | | MAE | MAPE | RMSE |
| Vehicle count | GPR | 13.37 | 11.86% | 22.23 |
| | Random forest | 20.68 | 15.97% | 33.36 |
| | GPR without additional spatial information | 29.49 | 27.72% | 45.71 |
| Fleet composition | GPR | 2.15% | 3.91% | 3.78% |
| | Random forest | 3.35% | 8.18% | 5.92% |
| | GPR without additional spatial information | 4.31% | 9.13% | 6.92% |

It is clear from Table 4 that the vehicle count per minute per grid and the fleet composition estimated by the proposed GPR model have smaller errors compared to the widely-used random forest model. We further observe that the additional geographical information (i.e., the POI information and OSM data utilized in this paper) can significantly improve the accuracy of the estimation.

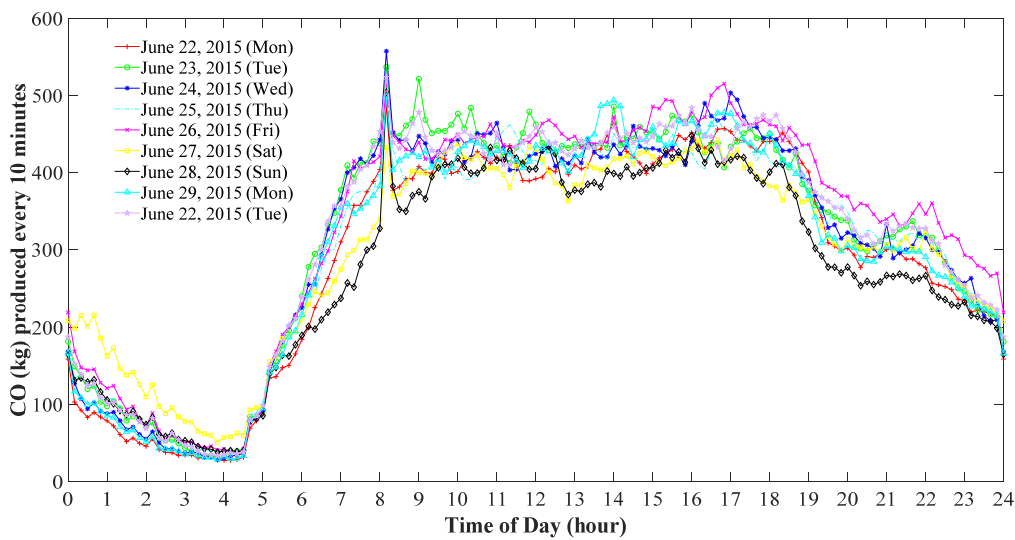

(a) CO emission

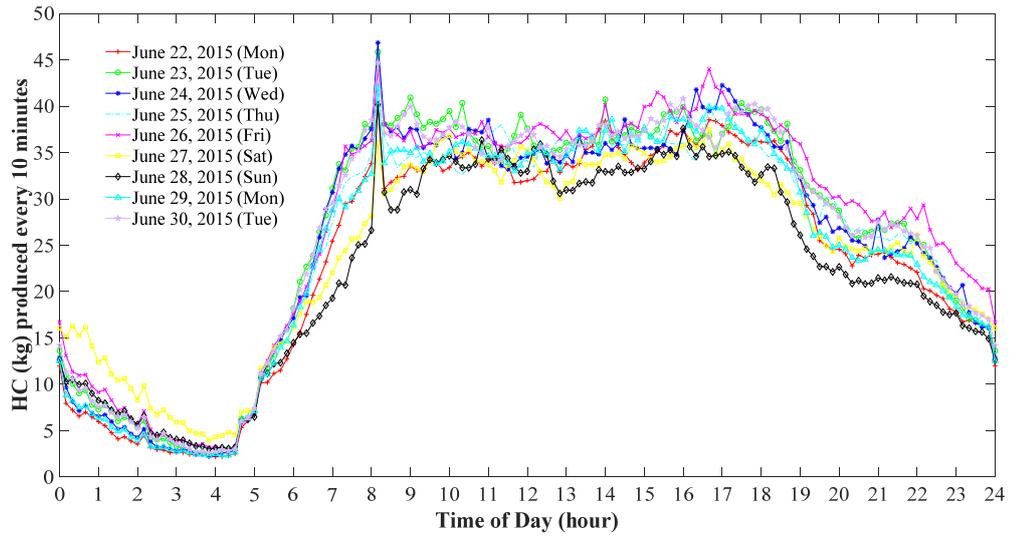

(b) HC emission

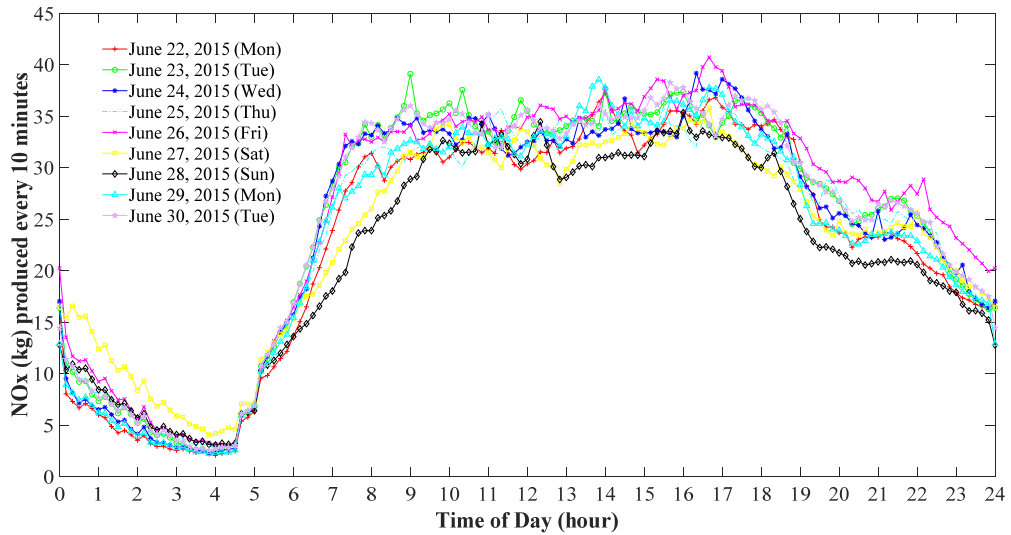

(c) NOx emission

Figure 8. Estimated emissions of the vehicle population for different pollutants.

Next, we estimate and analyze the emission patterns of the whole vehicle population. Figure 8 shows the temporal distributions of the three pollutants in the entire study area from the entire vehicle population. Notable differences from Figure 6 (taxi emissions only) are observed. First, we can clearly identify AM/PM peaks for the CO, HC, and $NO_x$ emissions of the entire vehicle population, unlike the case of taxis emission patterns. We can also clearly identify that AM peaks for weekends appear later than weekdays. This clearly shows that the proposed method is able to differentiate the temporal emission patterns of different vehicle types. Furthermore, during the daytime, taxis, which take up approximately 8% of the population (Figure 5), contribute to only 2.6% of the total traffic emissions. This is due to the much higher emission factors of heavy and other types of vehicles. Another interesting finding is that although the temporal patterns of these three pollutants are similar, the emission patterns at peak hours are a little bit different due to the differences in emission factors estimated in different grids.

Furthermore, the emissions produced during the AM peaks are commensurate with those during the PM peaks, despite that the PM peaks saw higher traffic volumes in the network than AM peaks (especially light vehicles; see Figure 5). This is due to (a) the surge of traffic in AM peaks creating heavy congestion and stop-and-go cycles that contribute to the higher emissions (causing the sharp increase of the emission factors); and (b) the fact that heavy vehicles have a major impact on emissions even with relatively low counts: During peak hours, heavy vehicles only take up 8% of the entire population but contribute approximately 20% of the total emissions; see Figure 9. On the other hand, during off-peak, heavy vehicles take up 10% of the population and contribute around 16% in 22:00-23:00. The significantly lower emission per heavy vehicle is likely due to the less congested road conditions during night times. These findings suggest the necessity of considering the temporal variations of vehicle volume and fleet composition, and that the taxi emissions cannot be extrapolated by static fleet information as is the case in existing studies. This justifies the contribution of our proposed method of emission extrapolation.

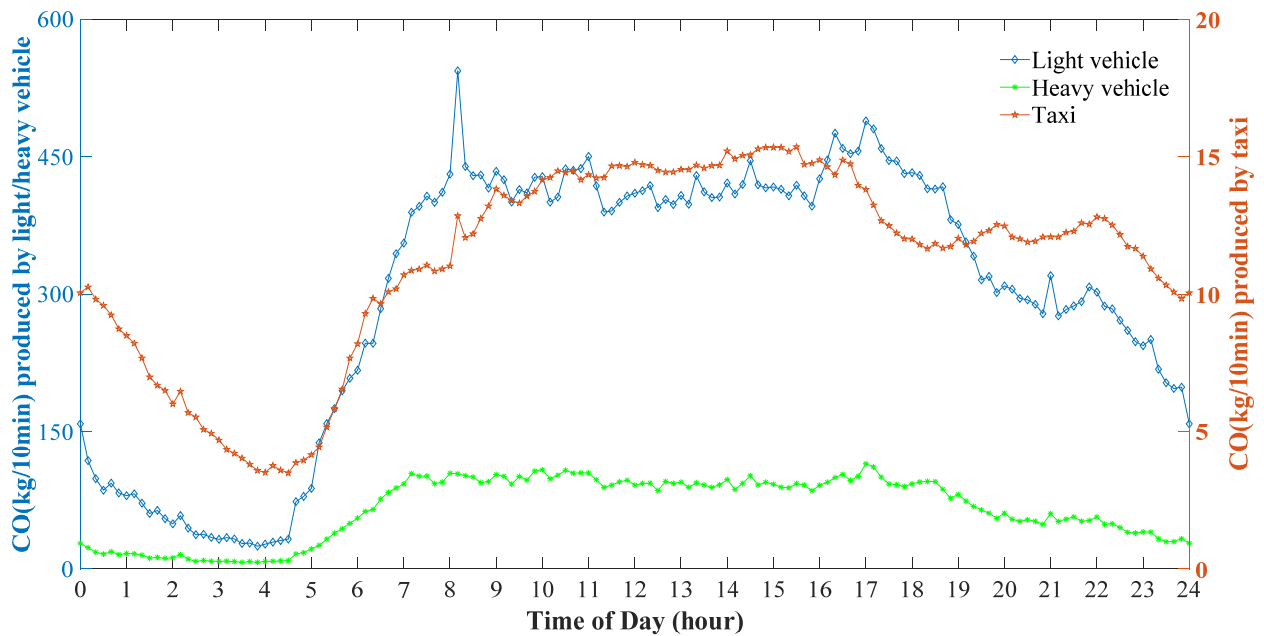

Figure 9. Estimated CO emissions of light vehicles, heavy vehicles, and taxi

Figure 10 shows the spatial patterns of the emissions (17:20-17:30). As shown in Figure 10(a), Hangzhou High-Speed Railway Station (1), Hangzhou Railway Station (2), Nanshan Road (3), Yanan Road (4), Desheng Highway Westbound (5), Zhonghe Highway Southbound (6), Qiushi Highway Southbound (7), intersection of Tianmushan Road and Zijingang Road (8) and Jiangnan Street (9) are identified as the areas with high CO emissions. Besides, Hangzhou Transit Terminal (10), Desheng Multidimensional Highway (11), Guashan Multidimensional Highway (12), Bus Terminal (13), Hang-Yong Expressway (14) and Xiasha Junction (15) from the airport to Hangzhou High-Speed Railway Station also have high CO emissions. This makes sense because Hangzhou High-Speed Railway Station (1), Hangzhou Railway Station (2), and Hangzhou Transit Terminal (10) are served as the main transportation terminals in Hangzhou, while Nanshan Road (3), Yanan Road (4), Desheng

Multidimensional Highway (11), Guashan Multidimensional Highway (12), Bus Terminal (13), and Hang-Yong Expressway (14) have heavy traffic flow in majority time. For the remaining areas, congestion and high traffic flow tend to occur in the rush hours.

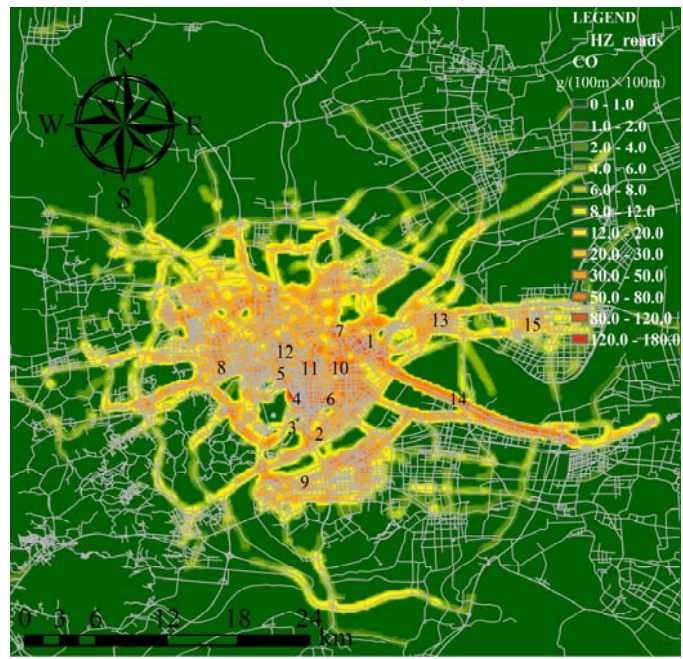

(a) Spatial distribution of CO emissions during 17:20-17:30.

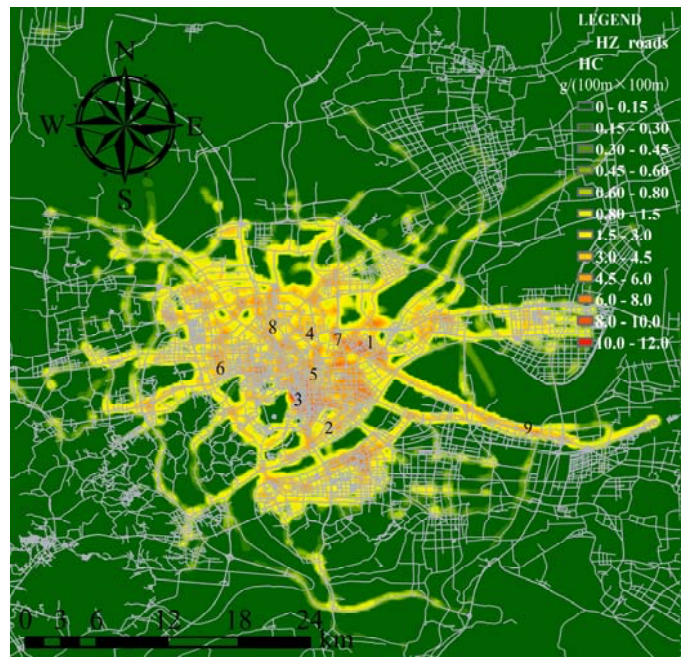

(b) Spatial distribution of HC emissions during 17:20-17:30.

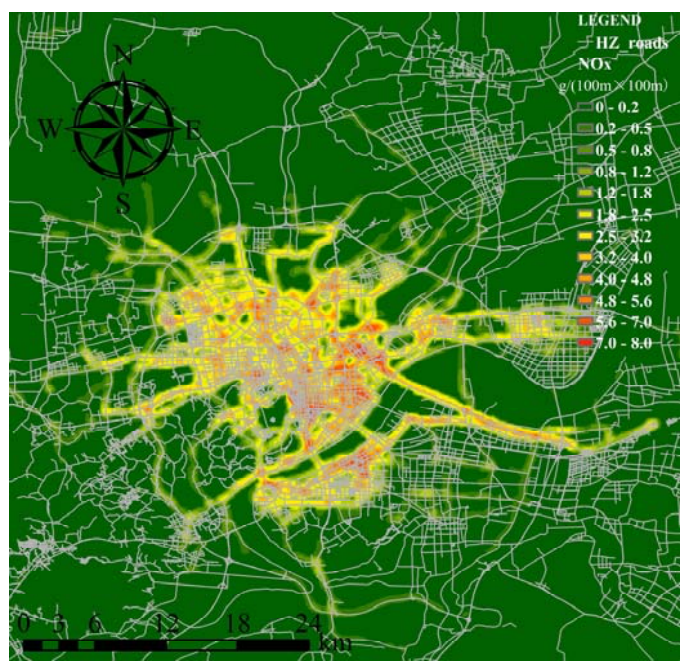

(c) Spatial distribution of NOx emissions during 17:20-17:30.

Figure 10. Spatial distribution of emissions for vehicle population.

As shown in Figure 10(b), areas with high HC emissions are similar to those with high CO emissions. Hangzhou High-Speed Railway Station (1), Hangzhou Railway Station (2), shopping districts besides the Westlake (3), Shixiang Road (4), Zhonghe Highway (5), the intersection of Tianmushan Road and Zijingang Road (6), Qiushi Highway Southbound (7), and the intersection of Wenyi Road and Jiaogong Road (8) have been identified.

Regarding the NOx emissions shown in Figure 10(c), it shares the similar spatial emission patterns with CO and HC. The main difference between them is that the amount of CO emissions are about ten times larger than those of HC and $NO_x$ emissions.

We further compare the proposed emission estimation method with two different methods: (a) one based on flow data from microwave detectors* and the static fleet information (provided by the Bureau of Statistics of Hangzhou), which is similar to the method proposed by Nyhan et al. (2016); (b) the other one based on the flow data and uniform dynamic fleet composition, both observed by LPR cameras. As shown in Figure 11, method (a) tends to underestimate the total emissions during the daytime, which is understandable given that the static fleet information may not be an accurate and up-to-date representation of the vehicle fleet during the study period. Moreover, method (b) tends to produce smooth emission profiles as a result of uniform extrapolation of the taxi emissions without accounting for spatial heterogeneity of the road network and traffic dynamics, while the proposed method captures these features and yield fine-granular temporal dynamics of the emissions as indicated by the local variations shown in the figure.

---

* The microwave data used in the case study are collected by 286 microwave detectors, which provide vehicle count, average speed and timestamp. There are 4,784,315 data entries within the study period.

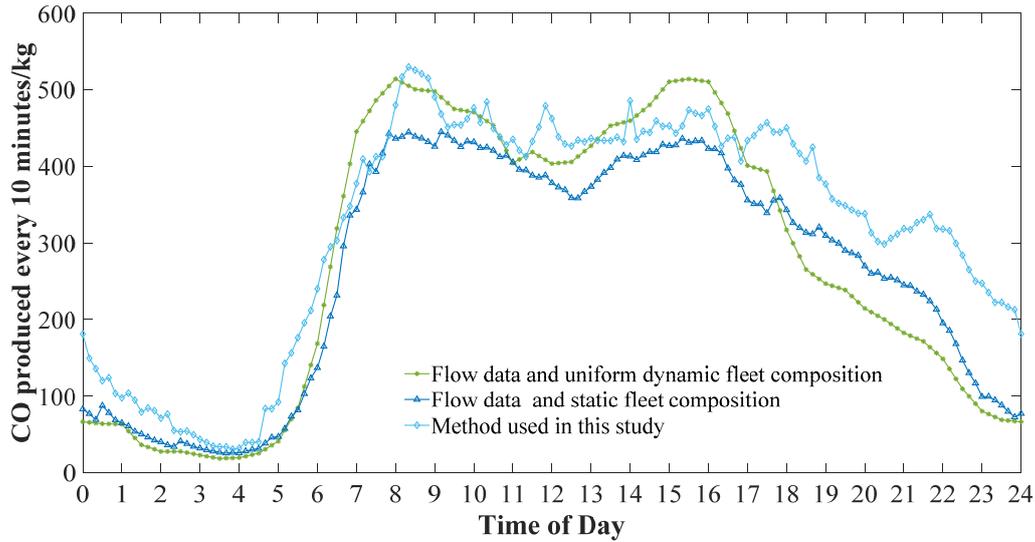

Figure 11. Comparison of estimated emissions with three methods.

## 4. CONCLUSIONS

This paper integrates taxi GPS data with multi-source urban data to reconstruct the spatial and temporal patterns of urban traffic emissions. An emission model is adopted to estimate emissions. The emission model enables to estimate emissions from different types of vehicles based on traffic patterns on road segments. After using this model to estimate the emissions from taxis, they are then mapped to spatial grids of urban areas to account for spatial heterogeneity. With GPR models, we estimate the spatially heterogeneous traffic volume and fleet composition and reveals high-resolution spatial-temporal patterns of traffic flows and emissions. The results of the case study show the vehicle emission in peak hours and non-peak hours is consistent with traffic flows, and identify emission hotspots in the city.

This research has several limitations mainly due to the datasets available. In order to reduce the influence of speed undulation on the emission factors, we calculate the non-zero speed records of taxis to estimate the average speed in each grid and this may underestimate the vehicle emissions in the case study. Another limitation is that we use the estimated number of vehicles in each grid due to the insufficient number of LPR cameras and their maldistribution, which may not provide the ground truth for a large-scale network. However, it should be noted that the spatial heterogeneity can be analyzed in a more precise way if a sufficient number of LPR cameras could be installed.

## ACKNOWLEDGMENTS

This research is financially supported by Zhejiang Provincial Natural Science Foundation of China [LR17E080002], National Natural Science Foundation of China [51508505, 71771198, 51338008], Key Research and Development Program of Zhejiang [2018C01007], and Fundamental Research Funds for the Central Universities [2017QNA4025].## REFERENCES

Aslam, J., Lim, S., Pan, X., Rus, D., 2012. City-Scale traffic estimation from a roving sensor network. In *Proceedings of the 10th ACM Conference on Embedded Network Sensor Systems*, November 6-9, 2012, Toronto, ON, Canada, pp. 141-154.21